# Consideration of resilience for digital farming systems

S. Bökle[1], L. Könn[1], D. Reiser[1], D.S. Paraforos[1], H.W. Griepentrog[1]
[1]Institute of Agricultural Engineering (440d), University of Hohenheim, Garbenstr. 9, 70599 Stuttgart, Germany
sebastian.boekle@uni-hohenheim.de

**Abstract**

Latest and current innovations of agricultural tech industry are increasingly driven by digital technologies. These digital farming solutions provide attractive advantages for farmers. The trend is going to devices and sensors, which send the acquired data directly to the cloud. Also the number of scientific publications on cloud based solutions follows this development. Considering on the other hand the necessity of continuous agricultural production in any kind of crises, new cloud-based digital systems and applications need to be reliable, independent of internet supply. In this conceptual study the necessary resilience is defined, which is marginally taken into account by agtech industry innovations. Problems of development using web-based farming systems are identified and discussed. For digital farming systems the farmers' individual needs of resilience are classified into five levels. Consequently, suggestions for soft- and hardware equipment are made. This includes the installation of a farm server, a local farm network, offline applications and consideration of edge computing, which can ensure a high level of resilience of new digital farming components.

**Keywords:** cloud, farm management information systems, internet blackout, resilience

**Introduction**

The more people that agriculture has to feed, the less it can afford any breakdowns within food production. Consequently, any technical solution that depends on a single, central system, needs to be avoided or developed fail-safe. While many farmers are still limited to their internet connection at the office area, working with desktop solutions, industry and science are increasingly focused on centralised online systems. Referring to this development, it's assumed for this article that internet connectivity is available also at farming area. A prevailing adaption of digital systems by farmers is also assumed. The used digital solutions by farmers need to provide continuous and fast functionality despite the system based vulnerabilities as shown in the next sections. Therefore independence of the actual landline connection, the provided mobile network or a provided bandwidth (Villa-Henriksen et al., 2020) would then mean increased resilience, which is the main focus of this paper.
Resilience describes the capacity to be able to absorb any possible influence and disturbance to retain the same essential function, structure, feedback and identity (Fröhlich-Gildhoff & Rönnau-Böse, 2019). On the one hand, resilience can be understood as purely external adaptability to the environment. Also, internal criteria can be taken into account (Hess, 2017). At the farm level, for example, resilience is the maximum difference between planned and actual yield at which the farm does not yet withdraw from production and the market. If the time span from the occurrence of the negative condition to the complete overcoming of the crisis on the one hand and resistance on the other is overstimulated, the resilience of the farm is considered as exhausted (Walker et al., 2004). Considering digital farming applications, resilience results in independence of power and internet supply.

FMISs (farm management information systems) or tools to develop prescription maps of recently introduced solutions are in most cases only accessible via online platforms. Only a few providers still offer desktop applications (Tummers et al., 2019). The aspect of data safety is barely considered or at least opaque to the available solutions. However, as these farm data fall into the category of proprietary company knowledge, they should also enjoy corresponding protection against abuse and malicious intents. Another aspect of resilience is interoperability. The lack of interoperability between devices, machinery, brands, soft- and hardware includes a high potential of dependency and consequently a conflict in terms of resilience. The potential of standardisation is insufficiently exploited and proprietary solutions are still dominant (Blank et al., 2013).

The aim of this work is to show the actual development of digitisation towards cloud-based solutions and its conflict in terms of resilience. Gradually ascending levels of hardware and software arrangements are suggested to effectively avoid insufficient resilience during internet blackouts. These levels ought to demonstrate what areas a farm needs to make resilient and where investments to do so should be made.

GNSS (global navigation satellite system) signals are of great importance in digital agriculture and also experience some vulnerability. But under several GNSS, which satellite signals are used simultaneously, it's very unlikely to have a black out of more than one at the same time. Therefore, in the present work the vulnerability of GNSS signals is not considered and assumed as stable. The most relevant constraint in this aspect is the decrease in accuracy from RTK (real-time kinematic) quality downwards. RTK signals, transferred by GSM (global system for mobile communications), suffer the corresponding limitations of internet-dependent applications and consequently need a redundant back up to ensure resilience.

**Potential threats on digital farming**

Development of cloud-based innovations

To reflect the increasing efforts in research about cloud-based solutions requests in the databases of SCOPUS and CABdirect with keyword pairs "agriculture" AND "cloud-based", - AND "server-based", - AND "remote server", - AND "remote storage" are illustrated in Figure 1 and show a clear increase in publications per year. The development starts with less than one publication per year before 2011 and reaches up to 53 in SCOPUS and up to 13 in CABdirect in 2019.

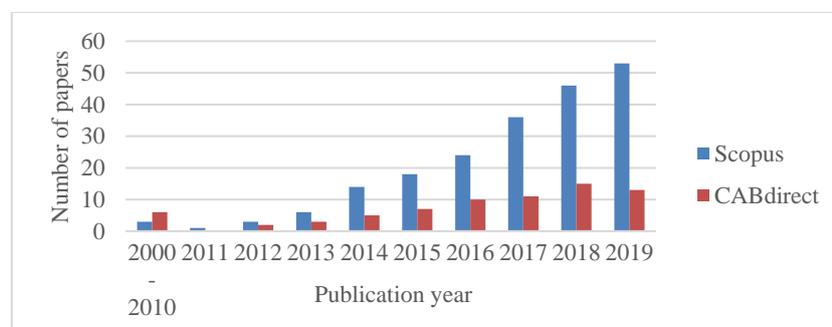

Figure 1. Number of publications between 2000 and 2019 concerning cloud-based innovations in agriculture at the databases of SCOPUS and CABdirect.

The FMIS plays an important role in digital farming systems (Paraforos et al., 2017). Tummers et al. (2019) identified in their SLR (systematic literature review), web-based FMISs as the biggest group within their results about FMIS types. Ströbel (2019) investigated 17 FMISs and at least seven have only cloud-based storage. Regarding the leading manufacturers, intensified cooperation in terms of cloud interoperability (Claver, 2019, 2020) can be observed. Efforts for offline solutions were more frequent in the past and are now receiving less and less attention in both academia and industry. Consequently, it should be acknowledged that farmers take a risk in relying on internet-dependent components of digital farming when using them for applications in critical time windows.

Internet, power and GNSS blackouts and cyber security

When past power and internet blackouts are considered, there is a large number of incidents with different origins revealing the vulnerability of internet and power supply. Environmental factors have a strong influence on the infrastructure. For example, in 2005, several regions in Germany were without electricity for up to six days as a result of extremely rough winter weather conditions (Schröder & Klaue, 2005). Accidents and a fire occurred in the building of a telecommunication provider that affected all services provided, including emergency calls for several hours (Reuter et al., 2020). In December 2015, there was a cyber-attack on the data network SCADA of a Ukrainian electricity provider that resulted in a blackout that affected over 200,000 people (Baezner, 2018). In 2016 the Cyber Division of the FBI published a warning that the agricultural sector might increasingly be targeted by cyber-attacks (FBI, 2016). Linsner et al. (2019) mentioned the increasing threat that technological vulnerabilities might be exploited by malicious agencies. The government in India is systematically shutting down internet access in various regions to prevent the organisation of riots via social media. At the same time, there is an enormous impact on all economic sectors of the country (Faleiro, 2020).

Problems of web-based solutions

Under normal, optimal conditions, farmers' digital farming components are in constant data communication via the internet, according to the mentioned assumptions. This allows all information needed for production management to be collected, exchanged, stored, processed and used for decision support. Figure 2 describes this state as Scenario I. All data carriers are connected via cloud services. These data consist of information such as weather, sensor, device and positioning data. Scenario II describes a worst-case situation. Here the farmer lacks any connection to the cloud. Consequently, as-applied information and yield documentation, sensors, weather data, telemetric services, management information and RTK data via GSM are not available. At the same time, there is no data exchange with the production equipment, facilities and devices of the farm itself, if they communicate via the cloud, like under normal conditions as described in the Scenario I, this results in a setback to bare offline functions. All data or work instructions must now be carried out via human-to-machine or human-to-human communication. Information beyond this is not available in Scenario II. Offline or desktop versions enjoy only a little attention in the farm industry and research, which leaves the farmer with limited stand-alone solutions.

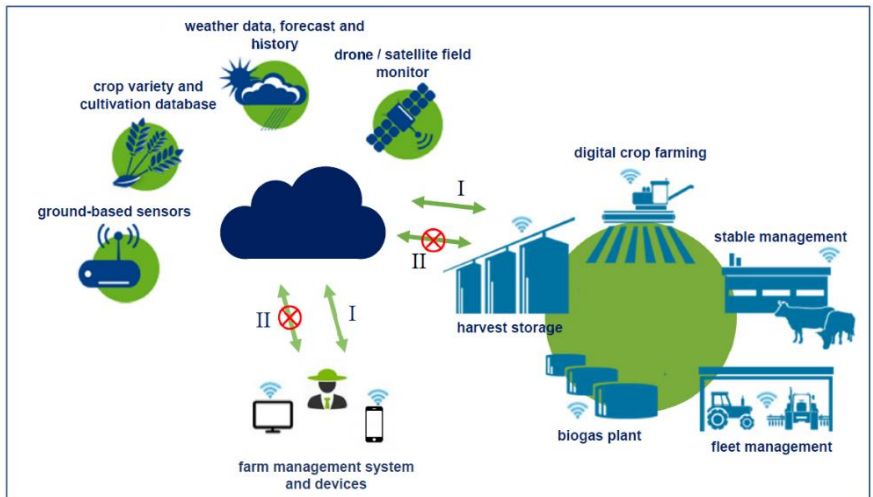

Figure 2. Connected farm: (I) Uninterrupted data transmission: FMIS, cloud, devices, and (II) Internet interruption, disconnections of all items (Baumbach, 2017).

**Integrating resilience in digital agriculture**

Different levels of resilience for hardware and software equipment
In general, there is a high heterogeneity among farms in terms of size, financial status, level of digitisation and supply of internet connection. Assuming there is internet connection and digital farm systems are adopted by farmers, this heterogeneity among farms result in different digitisation setups, fitting best to each farm business. This results in different expositions to different risks of blackouts and cyber-attacks and also variable implications associated with such disorders. Accordingly, farms have to build up resilience for significant functions of their digital system.

The authors propose a five-stage cascade to classify the resilience of digital farm systems. This was designed considering the different needs for resilience across farm business and to assist farmers and organisations in making the best decisions when choosing the best-balanced resilience arrangements. These five levels of resilience are presented in Table 1 and reflect increasing measures to gain resilience. For the levels in Table 1, an internet blackout is presupposed and that farmers do process all their data in the FMIS and not in an additional GIS (geographic information system). The vulnerability of GNSS signals won't be considered here, so it is assumed that GNSS signals are available and stable, excluded RTK corrections via GSM.

In the first level, no provisions for resilience are made, only functions independent of internet access remain functionable. As shown in Table 1, field boundaries, AB lines or management zones, which are stored in the tractor terminals, are available.

The second level includes a desktop application of farmers' FMIS, which allows some offline functions for decision support and other digital farming functions depending on the offline feature equipment of the FMIS. Data import and export is only possible via USB-Stick or Bluetooth.

To counter vulnerable centralised systems, the third level includes the farm server, which is a NAS (network-attached storage) server. It introduces decentralisation, which helps establish a more fail-safe operation. Necessary farm data are stored on the farm server. It also provides for more data safety and security as such hardware is a redundant storage system for specific data, which is also stored in the cloud as a back-up. The NAS server enables higher computing capacities with which more complex applications, mandatory for farm management, can be computed. This could include some forecasting software, which is not applicable on a usual farm computer. Furthermore, the management of access rights for third parties on the farm data stored in different clouds or on the farm server, could be conducted via the farm server. Another use would be the function as cache storage for sensors and devices with limited storage capacity for the acquired data. Like this possible data loss is prevented.

Table 1. Different stages of resilience considering digital farming features. X means the component is functional in case of internet interruption. (X) limited function.

| Components | Level of data resilience | | | | |
|---|---|---|---|---|---|
| | Level I | Level II (I + Desktop FMIS) | Level III (II + NAS-Server) | Level IV (III + LFN) | Level V (IV + AI) |
| GNSS Signals/ guidance/ positioning | x | x | x | x | x |
| Mobile RTK signal/guidance/ acurate positioning | (x) | (x) | (x) | x | x |
| Field boundaries | x | x | x | x | x |
| AB lines | x | x | x | x | x |
| Telemetry | | | | (x) | (x) |
| FMIS | | x | x | x | x |
| Farm server | | | x | x | x |
| LFN | | | | x | x |
| Documentation (Summary, Report) | | x | x | x | x |
| Documentation (As applied) | | x | x | x | x |
| Management zones (VRA) | x | x | x | x | x |
| Prescription maps (VRA) | | x | x | x | x |
| Yield mapping | | x | x | x | x |
| Soil mapping | | x | x | x | x |
| Irrigation management | | (x) | (x) | x | x |
| Proximal sensing (crop monitoring) | | x | x | x | x |
| RS (UAV) | | x | x | x | x |
| RS (satellite) | | | (x) last version manually | (x) last version manually | (x) last version updated by AI |
| Data Analysis (as applied documentation) | | | x | x | x |
| Data Analysis (e.g. of yield mapping) | | | x | x | x |
| Automatic data acquisition and transfer | | | | x | x |
| Weather data | | | | (x) data of own weather station | (x) data of own weather station |
| Forecasting (early warning systems) | | | | (x) data of own weather station | (x) data of own weather station |
| Public geodata | | | (x) last version manually | (x) last version manually | (x) last version updated by AI |
| Autonomous vehicles and robots | | | | (x) | x |
| AI | | | | | x |

The fourth level represents the maximum provisions for resilience purposes. The only limitations here are data actualisations from online sources, such as weather or satellite data and proprietary functions or applications of the OEMs (original equipment manufacturer). Compared to the previous level, level four is upgraded with a LFN (local farm network). Data between the farm server, sensors, machinery and devices are sent via this independent network. It includes signal solutions for positioning, maintaining controlled traffic and the transmission of coordinates of machinery and animals. In terms of section control and applications which need high and repeatable precision, RTK signals are replaced by the LFN. In this context, raw data are stored and partly processed on the farm server. AI (artificial intelligence) is not applied in level IV. Nevertheless, it is not intended to store all raw data on the farm server. This should mainly take place in the cloud where, for security purposes, decentralisation among different cloud servers should take place. When raw data has been processed, issues of interoperability (Villa-Henriksen et al., 2020) occur among FMIS, machinery, implements and sensors. Here open-source ontologies are recommended to make different data structures and formats readable for machines and humans. This requires that OEMs release their ontologies or adapt to international standard ontologies. Another suggestion for interoperability is edge computing as this can be realised on the sensor itself or on the NAS Server.

On the fifth level, artificial intelligence (AI) is added to the arrangements of the fourth level. The need for specific data from online sources gets identified and automatically updated. Of course, this proposition is also limited in case of an internet blackout. However, data could be gathered in advance before the farmer thinks of it, and the maximum extent of actualised data, until the internet cut off, is exploited. AI calculates for example relevant periods of needed satellite images according to the documented and planned crop rotation. Additionally, forecasts for pest management, as Chen et al. (2020) describe, could be run by AI on the farm server using data from the farm's own weather station.

**Conclusions**

With the increasing adoption of digital systems farmers must take into account that they underlie vulnerabilities which limit their reliability. Farmers face situations where they need redundant solutions in case digital tools fail because of unpredictable black outs. As every farm makes up different priorities in its farm business, farmers need to classify which level of resilience is needed, to maintain digitally supported food production during crisis without risking crucial reductions of product quality, process effectivity and environmental sustainability. Depending on the estimated frequency of blackouts and the impact of digital solutions on these three factors, investments into a higher level of resilience must be evaluated. Here level III reflects resilience in data availability and safety, level IV additional precise positioning and V computing capacities. Nevertheless, also with an increasing adoption of digital farming systems, the farmer stays the last resilient instance to ensure high quality food production

**Acknowledgements**

The present study was carried out within the MR digital project funded by the EIP-AGRI measure of the European Union. We thank the European Innovation Partnership "Productivity and Sustainability in Agriculture (EIP Agri)" for funding the project.